\documentstyle[epsf,preprint,eqsecnum,,aps]{revtex}
\begin{document}
\draft
\def\etal{{\it et al.}}
\def\ed{{\sl edited\ by\ }}
\def\vs{versus~}
\def\tj{$t$-$J$~}
\date{\today}
\title{Electronic Correlations\\
Near a Peierls-CDW Transition}
\author{P.~Monthoux\thanks{phm21@phy.cam.ac.uk}}
\address{Cavendish Lab, University of Cambridge\\ Madingley Rd, Cambridge
CB3 0HE, UK}
\author{D.J.~Scalapino\thanks{djs@vulcan.physics.ucsb.edu}}
\address{Department of Physics, University of California\\ Santa Barbara,
CA 93106-9530, US}

\maketitle

\begin{abstract}

Results of a phenomenological Monte Carlo calculation for a 2D
electron-phonon Holstein model near a Peierls-CDW transition are presented.
Here the zero Matsubara frequency part of the phonon action is dominant and
we approximated it by a phenomenological form that has an Ising-like
Peierls-CDW transition.  The resulting model is studied on a 32$\times$32
lattice.  The single particle spectral weight $A(k, \omega)$, the density
of states $N(\omega)$, and the real part of the conductivity
$\sigma_1(\omega)$ all show evidence of a pseudogap which develops in the
low-energy electronic degrees of freedom as the Peierls-CDW transition is
approached.
\end{abstract}

\newpage

\setlength{\baselineskip}{.3in}

\section{Introduction}

In the electron-phonon Holstein model, local phonon modes with frequency
$\omega_0$ are coupled to the electron charge density on each site of a
lattice.  Here we consider this model on a 2D square lattice with a
near-neighbor, one-electron transfer matrix element $t$. At half-filling
this system undergoes an Ising-like phase transition to a Peierls-CDW phase 
at a
transition temperature $T_c$. We are interested in developing a more
detailed picture of what happens to the electronic degrees of freedom as
this phase transition is approached.

When the temperature is reduced, the lattice site displacement-displacement
correlation length increases, diverging as $(T/T_c-1)^{-1}$ on an infinite
lattice.  This behavior clearly alters the usual Migdal, electron-phonon
renormalized, quasi-particle description 
\cite{Mig58}.  In particular, as
$T_c$ is approached, a pseudogap in what were the quasi-particle degrees of
freedom develops.  Previous studies of this type of
model have used various analytic approximations 
\cite{LRA73,Sad74,Mar90,MM00} 
as well as quantum Monte Carlo methods 
\cite{NSS91,VNW92}. However, the
non-gaussian nature of the lattice displacement fluctuations
\cite{Mon01,MM00} and their strong coupling to the electronic degrees of
freedom have raised questions \cite{MM00} about the analytic 
approximations
and the relatively small, 2D lattices 
$(10\times 10)$ that have been studied
with exact Monte Carlo algorithms have limited their range.  Here we
investigate this problem using a static, phenomenological Monte Carlo
approach, that allows us to treat the non-gaussian fluctuations and strong
coupling effects on larger lattices (32$\times$32) in the pseudogap
regime. In a static approximation, only the zero
frequency Matsubara part of the effective phonon
action is kept, which corresponds to treating the
thermal lattice fluctuations classically.  These
classical thermal fluctuations determine the
critical behavior at the Peierls-CDW transition and
the long-wave-length, low-frequency response of the
lattice. 

Previously, Monte Carlo simulations of classical
Heisenberg antiferromagnetic spins interacting with
electrons have been used to model the mangenites
\cite{BYM00} and recently a classical $xy$ action
has been used to study the effect of phase
fluctuations on a $d$-wave BCS system
\cite{LR01,Eck01}. Monien has studied the
cross-over from Gaussian to non-Gaussian behavior
for a 1D electronic system coupled to classical
incommensurate CDW fluctuations \cite{Mon01}. A
static field approximation in which the action was
obtained by integrating out the fermions has also
been used to treat the positive $U$ Hubbard model
\cite{BS00}. Here we will approximate the effective
static part of the phonon action using a
phenomenological Ising form.  We are interested in
the effect that an Ising-like Peierls-CDW
transition has on the electronic properties.
We find that the single particle spectral
weight $A(k,\omega)$, the density of states
$N(\omega)$, the spin susceptibility, and the
real part of the conductivity
$\sigma_1(\omega)$ all show evidence of a
pseudogap which develops in the low-energy
electronic degrees of freedom as the
Peierls-CDW transition is approached.

The Holstein Hamiltonian, which we will study,
consists of a one-electron hopping term $t$, local Einstein phonons of
frequency $\omega_0$ and an interaction which couples the on-site local
phonon displacement $x_i$ and the site charge density.
\begin{equation}
H=-t \sum_{\langle ij\rangle \sigma}
\left(c^\dag_{i\sigma}
c_{j\sigma} +
c^\dag_{j\sigma} c_{i\sigma}\right) + \sum_i\,
\left(\frac{P_i^2}{2M}
+ \frac{M\omega^2_0}{2}\ x^2_i\right) + g\
\sum_{i\sigma}
 x_i n_{i\sigma}
\label{one}
\end{equation}
The sum in the
first term is over the set of nearest neighbor
sites on a 2D square lattice. The 
$c^\dag_{i\sigma}$ fermion
operator creates an electron of spin $\sigma$ on site $i$, $P_i$ and $x_i$
are the conjugate momentum and position operators for the local
phonon mode on the $i^{\rm th}$ site, $g$ is the electron-phonon coupling,
and $n_{i\sigma} = c^\dag_{i\sigma}c_{i\sigma}$. 

In the usual quantum Monte Carlo approach, the fermions are traced over
leaving an expression for the partition function in terms of an effective
action $S[\{x_i(\tau)\}]$ for the lattice displacement field $x_i(\tau)$.
Here we imagine that this has been done and then express $x_i(\tau)$ in
terms of its Matsubara frequency transform.
\begin{equation}
x_i(\tau)= \sum_m\, x_i (\omega_m)\, e^{-i\omega_m\tau}
\label{two}
\end{equation}
with $(\omega_m=2m\pi T)$. 
As the transition temperature is approached, the electron-phonon coupling
leads to a softening of the phonon frequencies and 
for $Q=(\pi, \pi)$, 
$\omega_Q$ vanishes at $T_c$. In the region near
$T_c$ where $\omega_Q \ll
2\pi T$, the effective action is dominated by the zero frequency
Matsubara part. As discused by Bickers and one
of the authors \cite{BS00}
for the Hubbard model, one could now proceed to directly calculate the
$\omega_m=0$ contribution to the action by integrating out the fermions in the
presence of a given static field configuration. Here, however, we will 
approximate the
effective action by an Ising-like form
\begin{equation}
S[\{x_i\}] = \beta\left(\frac{b}{4}\ \sum_i\, \left(x^2_i-a^2\right)^2 + c
\, \sum_{\langle i_{ij}\rangle}\, x_i x_j\right)
\label{three}
\end{equation}
with $x_i\equiv x_i(\omega_m=0)$. Here, $a$, $b$, and $c$ vary slowly with
temperature and we will treat them as phenomenological constants.  In the
following we will set the scale of $x_i$ so that $a=1$ and take $b=2$ and
$c=1/8$ in units where the hopping $t=1$.
These values imply $\langle
x^2_i\rangle \simeq 1$ and give a mean field transition temperature
$T_c^{MF}=0.5t$. 

Now the calculation of the electronic properties reduces
to the problem of electrons moving in a static random external field
$\{x_i\}$ with
\begin{equation}
H=-t\, \sum_{\langle i, j\rangle\sigma} 
\left(c^\dag_{i\sigma}c_{j\sigma} +
h.c.\right) + g\, \sum_{i\sigma} x_i n_{i\sigma}\ .
\label{four}
\end{equation}
The probability distribution of the external field is proportional to
\begin{equation}
P[\{x_i\}] = \frac{\exp (-S[\{x_i\}])}{Z}
\label{five}
\end{equation}
with $S[\{x_i\}]$ given by eq.~(\ref{three}) and 
\begin{equation}
Z=\int\, \left(\prod_i dx_i\right)\, \exp(-S[\{x_i\}])\ .
\label{six}
\end{equation}
In the following we will set the coupling $g$ in eq.~(\ref{four}) equal to 1.

Even on a 32 by 32 lattice, certain physical quantities such as the density
of states show significant size effects due to the degeneracies of the simple
nearest-neighbor tight-binding spectrum we are using. In order to break these 
degeneracies and obtain results closer to the continuum limit, we follow 
Assaad \cite{Ass01} and introduce a small magnetic 
field that vanishes in the 
thermodynamic limit. The magnetic field is introduced via the Peierls 
phase factors and the electronic Hamiltonian becomes

\begin{eqnarray}
H & = &\, \sum_{\langle i, j\rangle\sigma} 
\left(c^\dag_{i\sigma}
t_{ij} c_{j\sigma} +
h.c.\right) + g\, \sum_{i\sigma} x_i n_{i\sigma} = \sum_{\langle
ij\rangle\sigma} c^\dag_{i\sigma} H_{ij} [\{x\}]
\, c_{i\sigma}\\ 
\noalign{\hbox{with}}
t_{ij} & = & -t e^{i{2\pi\over\Phi_0} \int_i^j {\bf A}(l)\cdot d{\bf l} }\
\label{HL}
\end{eqnarray}

\noindent where ${\bf B} = \nabla \times {\bf A}$ and $\Phi_0$ is 
the flux quantum. The size of the magnetic field 
is $B = \Phi_0/N$ where $N=L\times L (L=32)$
is the number of lattice sites. We use a
symmetric gauge with 
\begin{equation}
\vec A (\ell) = \frac{B}{2} (-\ell_y, \ell_x, 0)
\label{onenine}
\end{equation}
and boundary conditions such that
\begin{eqnarray}
c^\dag_{\vec\ell+L\hat x} &=&
e^{i\frac{\pi}{L}\, \ell_y}\,
c^\dag_{\vec\ell}\nonumber\\
c^\dag_{\vec\ell+L\hat y} &=&
e^{-i\frac{\pi}{L}\, \ell_x}\, c^\dag_{\vec
\ell}
\label{oneten}
\end{eqnarray}
As Assaad showed, the introduction of such a
magnetic
field removes the level degeneracy spreading
the single particle states over the bandwidth.
This procedure provides a significant
reduction in the finite size effects and we
will use it for all the quantities that will
be calculated.  While it breaks translational
invariance for a finite lattice size $L$, the
invariance is naturally restored when $L$ goes
to infinity. On our finite lattice, we will
simply take the average over all lattice sites
of the quantities of interest.

\section{Calculations}

We are interested in the fermion spectral function and density
of states which can be obtained from the single particle Green's
function:
\begin{equation}
G_{ij\sigma}(\tau,\tau') =
-\langle\langle T_\tau\{c_{i\sigma}(\tau)
c^\dag_{j\sigma}(\tau'
)\}\rangle_f\rangle_{\{x\}}
\label{GreenFunc}
\end{equation}
The double average $\langle\langle
\cdots\rangle_f\rangle_{\{x\}}$ is defined as
\begin{eqnarray}
\langle\cdots\rangle_f & \equiv & 
Tr[\exp(-\beta H)\cdots]\over Tr[\exp(-\beta H)]
\label{AverageF} \\
\langle\cdots\rangle_{\{x\}} & 
\equiv & {1\over Z}\int \Big(\prod_i
dx_i\Big)\exp(-S[\{x\}
])\cdots
\label{AverageX}
\end{eqnarray}
where $H$, $S[\{x\}]$ and $Z$ are given by
eqs.~(\ref{four}), (\ref{three}), (\ref{six}) respectively.
In order to carry out the fermion trace it is convenient to make the
canonical transformation
$\{c^\dag_{i\sigma},c_{i\sigma}\}\longrightarrow
\{\psi^
\dag_{i\sigma},\psi_{i\sigma}\}$ that diagonalizes the Hamiltonian
matrix
$H_{ij}[\{x\}]$, eq.~(\ref{four}).
\begin{eqnarray}
c_{i\sigma} = \sum_{r}O_{ir}[\{x\}]\psi_{r\sigma} \quad ; \quad
c^\dag_{i\sigma} & = & \sum_{r}O^*_{ir}[\{x\}]\psi^\dag_{r\sigma}
\label{Canon} \\
\sum_{ij} O^*_{is}[\{x\}] H_{ij}[\{x\}] O_{ir}[\{x\}] & = &
\epsilon_r[\{x\}]\delta_{rs}
\label{Eigenvalues} \\
\sum_{ij} O^*_{is}[\{x\}]O_{ir}[\{x\}] & = & \delta_{rs}
\label{Ortho}
\end{eqnarray}

After carrying out the fermion trace, and analytically continuing to real
frequencies, the retarded single particle Green's 
function is obtained as
\begin{eqnarray}
G^R_{ij\sigma}(\omega) = \Big<\sum_r O_{ir}[\{x\}]O^*_{jr}[\{x\}]{1\over
\omega -\epsilon_r[\{x\}] + i0^+}\Big>_{\{x\}}
\label{GreenR}
\end{eqnarray}
While averaging over the Monte Carlo
configurations $\{x\}$ restores the
translational symmetry breaking associated
with the individual displacement field
configuration, the introduction of the
magnetic field, eq.~(\ref{HL}), breaks
translational symmetry.  As noted, we are
willing to accomodate this feature in order to
reduce the finite size effects that appear in
the frequency dependence of $A(k,\omega)$ and
$N(\omega)$. With this in mind, we define
the Fourier transform
of the single particle Green's function as:

\begin{equation}
G^R({\bf k},\omega) = {1\over N}\sum_{ij}e^{i{\bf k}({\bf R}_i-{\bf
R}_j)}G^R_
{ij\sigma}(\omega)
\label{FourierT}
\end{equation}
where $N$ is the number of sites. The spectral function $A({\bf
k},\omega)$  is then given by:

\begin{eqnarray}
A({\bf k},\omega) & = & -{1\over \pi} {\rm Im} G^R({\bf k},\omega)
\nonumber\\ 
& = & \Big<{1\over N}\sum_{ij}e^{i{\bf k}({\bf R}_i-{\bf R}_j)}\sum_r
O_{ir}[\{x\}]
O^*_{jr}[\{x\}]\delta(\omega -\epsilon_r[\{x\}])\Big>_{\{x\}}
\label{SpecFunc}
\end{eqnarray}
and the single-particle density of states $N(\omega)$ is

\begin{eqnarray}
N(\omega) & = & -{1\over \pi}\, \frac{1}{N}
\, \sum_{\bf k} {\rm Im} G^R({\bf k},\omega)
\nonumber \\
& = & \left\langle\frac{1}{N}\, \sum_{ir} 
|O_{ir}[\{x\}]|^2\delta
(\omega
-\epsilon_r[\{x\}])\right\rangle_{\{x\}}
\label{DOS}
\end{eqnarray}
In the numerical calculations, the
$\delta$-functions are then replaced by
Lorentians of width $\eta = 0.01t$:
\begin{equation}
\delta(\omega_i-\epsilon_r) \rightarrow {1\over \pi} {\eta\over 
(\omega_i-\epsilon_r)^2 + \eta^2}
\label{delta}
\end{equation}

We are also interested in the real part of the 
conductivity
$\sigma_1(\omega)$ which can be obtained from the analytic continuation of
the usual current-current correlation function
\begin{equation}
D_{ij}(\tau) = \langle\langle T_\tau \left\{J^x_i (\tau) 
J^x_j (0)\right\}\rangle_f\rangle_{\{x\}}\ .
\label{twenty}
\end{equation}
Here,
\begin{equation}
J^x_i = i\, \sum_\sigma\,
\left(t_{i+\hat x i}c^\dag_{i+\hat x\sigma}c_{i\sigma}
-t_{i i+\hat x}c^\dag_{i\sigma}c_{i+\hat
x\sigma}\right)
\label{twentyone}
\end{equation}%
with $\hat x$ a unit displacement in the $x$-direction.
Analytically continuing the Matsubara frequency transform of
$D_{ij}(\tau)$ to real frequencies, we obtain the retarded propagator
whose imaginary part is given by
\begin{eqnarray}
{\rm Im} D^R_{ij}(\omega) & = & 2 {\rm Im}\left\langle\sum_{rs} 
M_{ij;rs}[\{x\}]{f(\epsilon_r) - f(\epsilon_s)\over \omega - \epsilon_r[\{x\}] +
\epsilon_s[\{x\}] + i 0^+}\right\rangle_{\{x\}} \label{JJ2} \\
\noalign{\hbox{with}}
M_{ij:rs}[\{x\}] & = &m_{i;rs}m^*_{j;rs} + m_{j;sr}m^*_{i;sr}
- m_{i;rs}m_{j;sr} - m^*_{j;rs}m^*_{i;sr} \\
\noalign{\hbox{and}}
m_{i;rs} & = & t_{ii+\hat x}O_{i+\hat x r}[\{x\}]O^*_{is}[\{x\}]
\label{MatEl}
\end{eqnarray}
The optical conductivity $\sigma_1(\omega)$ is 
then obtained from the imaginary part of the
retarded current-current correlation function ${\rm Im}D^R_{ij}(\omega)$, 
eq.~(\ref{JJ2}) as

\begin{equation}
\sigma_1(\omega) = {1\over{N\omega}} \sum_{ij} {\rm Im} D^R_{ij}(\omega)
\label{Sigma1}
\end{equation}

A check on the calculation of $\sigma_1(\omega)$ is provided by the f-sum
rule relating the average kinetic energy per site in the $x$-direction
to the integral over all frequencies of $\sigma_1(\omega)$. 
The kinetic energy operator $K^x_i$ for site i is

\begin{equation}
K^x_i = \sum_\sigma
\left(t_{i+\hat xi}c^\dag_{i+\hat x\sigma}c_{i\sigma}
+t_{ii+\hat x}c^\dag_{i\sigma}c_{i+\hat x\sigma}\right)
\label{Kinetic}
\end{equation}
Its average is readily obtained as
\begin{equation}
\left\langle\left\langle K^x_i\right\rangle_f\right\rangle_{\{x\}} 
= \left\langle\sum_r {\rm Re} \left(t_{ii+\hat x}
O_{ir}[\{x\}]O^*_{i+\hat xr}[\{x\}]f(\epsilon_r)\right)\right\rangle_{\{x\}}
\label{KineticAv}
\end{equation}
leading to the f-sum rule
\begin{equation}
\int_0^\infty d\omega\sigma_1(\omega) = -{\pi \over 2}<<K^x_i>_f>_{\{x\}}
\end{equation}

In the same manner, one can study the magnetic
spin susceptibility
\begin{equation}
\chi(q,\tau) = \langle\langle T\{M_z(q,\tau)
M_z(q,0)\}\rangle_f\rangle_x
\label{twotwentyone}
\end{equation}
with
\begin{equation}
M_z(q) = \sum_\ell e^{iq\cdot\ell}
\left(\frac{n_{\ell\uparrow}-n_{\ell\downarrow}}{2}\right)
\label{twotwentytwo}
\end{equation}
Here we are interested in the $q$ goes to zero
static Pauli spin susceptibility which is
given by
\begin{equation}
\chi_0 = -\left\langle{2\over N}\sum_{ijrs}O_{ir}[\{x\}]O^*_{is}[\{x\}]O_{js}[\{x\}]
O^*_{jr}[\{x\}]{f(\epsilon_r) - f(\epsilon_s)\over \epsilon_r[\{x\}] -
\epsilon_s[\{x\}]}\right\rangle_{\{x\}}
\label{twotwentythree}
\end{equation}

The calculations were carried out on a 32$\times$32 lattice. 
The probability distribution $P[\{x\}]$, eq.~(\ref{five}), is sampled 
with a Hybrid Monte Carlo algorithm \cite{SSS86,Duane,Creutz}, 
which we have found to be more efficient than the local Metropolis 
updating scheme. A step of the Hybrid Monte Carlo algorithm consists of a 
proposed move and an accept/reject procedure. A move is proposed by 
generating Gaussian distributed auxiliary momenta with zero mean 
and variance equal to the temperature. Newton's equations 
of motion using $S[\{x_i\}]/\beta$, eq. ~(\ref{three}), as the potential 
energy are integrated for n steps with a time-reversible integration scheme 
(necessary for detailed balance), such as the Verlet algorithm that we used. 
The proposed move is accepted with probability 
$min(1,\exp(-\beta \Delta E))$ where $\Delta E$ is the difference in energy 
between the final and initial configuration and the energy is the sum of the 
potential energy $S[\{x_i\}]/\beta$ and the kinetic energy of the 
auxiliary momenta. The accept/reject step corrects for the approximate 
integration of Newton's equations of motion and yields an exact algorithm.
The number of molecular dynamics steps n and the time step used for 
the integration must be chosen to maximize the efficiency of the 
sampling procedure. We set these parameters by maximimizing the average 
distance between an initial configuration and the configuration after one 
step of the Hybrid Monte Carlo procedure. We found that 20 molecular 
dynamics steps with a time step of 0.1 gave a consistently good 
performance for all of the temperatures of interest. 
For these values of the parameters, the acceptance ratio, which 
depends on how well the energy is conserved by the approximate integration
of Newton's equations, turns out to be nearly independent of temperature 
and between 85\% and 90\% for the $32\times 32$ lattice. Note that this 
acceptance ratio does depend on the size of the system, because the 
energy is an extensive quantity, and decreases as the system size 
increases.

Other parameters of the algorithm that must be set are the number of 
thermalization steps taken before any measurements and how many steps of
the algorithm are necessary to generate statistically independent 
measurements. While these parameters can be different for different 
observables, we have looked at the average value of the action,
eq. ~(\ref{three}), and the order parameter $x_s = {1\over N}\sum_i(-1)^ix_i$ 
which we expect to be the slowest observable to thermalize and 
decorrelate. Since for a finite system one can sample configurations 
with both $x_s > 0$ and $x_s < 0$ during a run, one must 
actually monitor the absolute value of $x_s$ in the simulation. 
Looking at the Monte Carlo time series of values of $|x_s|$ and 
$S[\{x_i\}]$, we found that 10,000 thermalization steps of the 
Hybrid Monte Carlo algorithm were adequate to equilibrate the 
system for our $32\times 32$ lattice. 

The number of steps of the sampling procedure necessary to generate
statistically independent measurements was determined by the standard
data blocking procedure. The values of the observables are averaged over
1,2,3,...n consecutive Hybrid Monte Carlo steps and the variance computed. 
When the number of steps n is greater than the correlation time of the 
sampling procedure, the variance no longer changes. A plot of the 
variance versus n allows one to determine the number of steps needed to 
generate statistically independent measurements easily. For all 
temperatures of interest, 100 steps of the Hybrid Monte Carlo 
algorithm were adequate to decorrelate the values of $|x_s|$ 
and $S[\{x_i\}]$.

Note that the parameters of the algorithm need not be the same for 
each temperature. The correlation and thermalization times are smaller 
at high temperatures. But given that by far the most time consuming 
part of the simulations is the calculation of the electronic
observables, any fine tuning of the Hybrid Monte Carlo sampling 
procedure would only lead to marginally small gains in computer time.

We carried out simulations at 11 temperatures starting with 0.5t 
progressively reducing the temperature down to 0.25t. 
Except for the highest temperature, the initial configuration was 
taken as an equilibrium configuration of the previous simulation.
Each of the runs consisted of 1024 sample configurations with 100 
Hybrid MC steps between successive measurements. The calculations were 
carried out on 64 processors of an IBM SP parallel computer, using a
different seed of the random number generator on each processor.
Thus 16 sample configurations from each processor were needed and 
the total length of a run on a given node was 1600 Hybrid Monte Carlo
steps. Close to and below $T_c$, this is not sufficiently long for 
the system to sample configurations with positive and negative values 
of $x_s$ during a run and thus the system is trapped near one of 
the two peaks ($x_s >0$ or $x_s < 0$) of the probability 
distribution $P[\{x\}]$, eq.~(\ref{five}). 
Since one expects the system to end up in a state with $x_s >0$ or
$x_s < 0$ with equal probability as the temperature is lowered, on average
one ought to sample configurations with $x_s > 0$ on half of the processors 
and configurations with $x_s < 0$ on the other half.

The data analysis was carried out with the multiple histogram 
technique of Ferrenberg and Swendsen \cite{ferrenberg}. 
This technique uses the information available from all of the runs
and allows one to obtain values for the observables for any 
temperature in the range covered by the simulations. But it requires
the range of energies sampled from simulations at neighboring
temperatures to overlap and this determined the number of runs (11) that 
we carried out. The statistical error estimates were obtained
with the bootstrap method. Our simulation data was resampled 128 times,
which allows an estimation of the error bars to a bit better that 10\%.
The multiple histogram calculation of the partition functions and 
observables must be repeated for each bootstrap sample and the 
statistical errors obtained as the variance of the 128 multiple 
histogram values of the observables. With 1024 samples, the 
statistical errors were typically of the order of 1\% in the spectral 
function $A(k,\omega)$ and 0.3\% in the density of states $N(\omega)$ 
and the optical conductivity $\sigma_1(\omega)$.

\section{Results}

For the parameters $a=1, b=2$, and $c=1/8$ that we have chosen, the mean
field transition temperature $T_c^{MF}=0.5t$. Results for the
Ising-specific heat on a 32$\times$32 lattice are plotted in Fig.~1 and
imply that the actual transition temperature is close to $T_c=0.27t$. As
discussed, we are interested in the temperature 
region $T_c^{MF}>T>T_c$ and particularly the
behavior as $T$ approaches $T_c$ where
the $\omega_m=0$ Matsubara frequency part of the action is dominant.
In this region, the Peierls-CDW rms gap $\Delta_{rms}=g\sqrt{\langle
x^2\rangle}$ is of order $t$ for the parameters
we have selected. Thus, the basic coherence 
length, which
varies as $t/\Delta_{rms}$ is of order the lattice spacing and the
important length scales are the Peierls-CDW correlation length
$\xi_{PCDW}(T)$ and the quasi-particle thermal correlation length
$\xi_{th}(T)$. For a non-interacting system, the thermal quasi-particle length 
for ${\bf k}\simeq (\pi/2, \pi/2)$ is $\xi_{th}(T)
\simeq 2t/\pi T$, while for ${\bf k}\simeq (\pi,0)$,
$\xi_{th}(T)$ varies as $2\sqrt{t/T}/\pi$ due to the
Van Hove singularity.

Results for $\xi_{th}(T)$ for ${\bf k}=(\pi/2, \pi/2)$ and
${\bf k}=(\pi, 0)$ are shown as the long- and short-dashed lines
respectively in Fig.~2. The temperature dependence of the
Peierls-CDW correlation length $\xi_{PCDW}(T)$ obtained from a Monte Carlo
simulation of $\langle x_ix_j\rangle$ using the action given by
eq.~(\ref{three}) is also plotted in Fig.~2
\cite{note}. Here, one sees that as the
temperature is lowered from $T^{MF}_c$ towards $T_c$, the Peierls-CDW
correlation length exceeds $\xi_{th}(T)$ as $T_c$ is approached.  
We also see that this occurs at a higher temperature for 
quasi-particles moving in the [1,0] direction
(${\bf k_F}=(\pi, 0)$) than for quasi-particles moving in the [1,1] 
direction (${\bf k_F}=(\pi/2, \pi/2)$) as expected.

Turning now to the electronic properties, we first look at the single
particle density of states $N(\omega)$ given by eq.~(\ref{DOS}).
Results for $N(\omega, T)$ \vs $\omega$ for various values of the
temperature are shown in Fig.~3 and a
three-dimensional plot of $N(\omega, T)$ versus
$\omega$ and $T$ is shown in Fig.~4. 
The lattice fluctuations at the higher
temperatures have wiped out the Van Hove peak.
As the temperature is decreased and the
Peierls-CDW correlation length exceeds the
quasi-particle thermal correlation length,
coherence peaks appear in $N(\omega, T)$.
This behavior is similar to what has been
found for the 2D Hubbard model near half-filling
as $T$
goes towards zero and the antiferromagnetic
correlation length exceeds the thermal
quasi-particle length \cite{VT96,Mou00}.

In Figs 5a and 5b we show the single-particle
spectral weights $A(k,\omega,T)$ for
${\bf k} =(\pi/2, \pi/2)$ and $(\pi,0)$ respectively.
Again, at higher temperatures, the thermal
lattice fluctuations give rise to a broad peak
in these spectral weights, due to the static
random nature of the lattice displacements.
As the temperature is lowered and the lattice
displacement correlation length increases,
coherent quasi-particle peaks, associated with
the Peierls-CDW phase which onsets at $T_c$
are clearly seen.  A careful examination shows
that these coherence peaks appear at a higher
temperature for ${\bf k} =(\pi,0)$ than for
${\bf k} =(\pi/2, \pi/2)$. Thus, the Peierls-CDW gap
appears to open first at the ${\bf k}=(\pi,0)$
region of the Fermi surface and then at a
lower temperature for the diagonal ${\bf k} =(\pi/2,
\pi/2)$ region. This simply reflects the fact
that the Peierls-CDW correlation length first
exceeds the ${\bf k} =(\pi,0)$ quasi-particle thermal
correlation length, as seen in Fig.~2.

In Fig.~7, we plot the static spin
susceptibility $\chi_0(T)$ versus $T$. As the
temperature decreases towards $T_c$,
$\chi_0(T)$ clearly shows the development of a
pseudogap.
Finally, turning to the real part of the conductivity, we have plotted
$\sigma_1(\omega)$ \vs $\omega$ and $T$ in
Fig.~7. As noted, the area under $\sigma_1(\omega)$ should be 
equal to $-\pi/2$ times the $x$-kinetic energy per site. 
If the integral of $\sigma_1(\omega)$ over $\omega$ is performed 
analytically for each Monte Carlo configuration, this sum
rule is satisfied to within statistical errors. 
The thermally randomized lattice gives rise to elastic scattering at higher
temperatures. As the temperature is lowered, the pseudogap
in the quasi-particle spectrum gives rise to a shift in
$\sigma_1(\omega)$ spectral weight from low frequencies to frequencies
above $2\Delta_{rms}$.

\section{Conclusion}

The electronic properties of a 2D electron-phonon
model which undergoes a commensurate Peierls-CDW
transition have been studied within a static
approximation which takes into account only the
$\omega_m=0$ Matsubara fluctuations of the lattice.
Using a Monte Carlo calculation based upon a
phenomenological Ising-like form for the lattice
displacements, the effect of these classical thermal
lattice fluctuations on the one-electron density of
states, the single particle spectral weight, the
spin susceptibility and the optical conductivity
were determined.  Each of these quantities shows
evidence of a depletion of low-energy electronic
degrees of freedom as a pseudogap opens.

For the parameters which we have chosen, $\xi_0\sim
t/\Delta_{\rm rms}$ is of order the lattice spacing
so that the important lengths for this 2D
lattice are the thermal
quasi-particle length and the Peierls-CDW
correlation length \cite{VT96,Mou00}.  
As the temperature is lowered
toward $T_c$, the Peierls-CDW correlation length
first exceeds the thermal quasi-particle correlation
length for electrons near $\vec k\sim (\pi, 0)$ and
a pseudogap begins to open over the parts of the
fermi surface associated with the Van Hove
singularity of the non-interacting system.  Then as
the temperature is further decreased, a temperature
is reached such that $\xi_{\rm PCDW}(T)$ exceeds the
thermal quasi-particle coherence length for the
diagonal $(\pi/2, \pi/2)$ region and a pseudogap is
opened over the entire fermi surface.

This continuous depletion of the single particle
density of states at the fermi energy appears as a
suppression of the Pauli spin susceptibility.  In
addition, spectral weight in the optical
conductivity is shifted from lower frequencies to
higher frequencies, eventually appearing as a peak
above $\Delta_0$, resembling $\sigma_1(\omega)$ in
the ordered Peierls-CDW phase.

While the results we have presented are in
agreement with various
theoretical and numerical expectations, we
believe that this numerical
method which takes into account the $\omega_m=0$
Matsubara fluctuations
represents a useful approach to a more detailed
understanding of the electronic properties of a
2D electron-phonon system as a commensurate
second order phase transition is approached.
From a diagramatic
point of view, it sums all Feynman graphs
associated with $\omega_m=0$
phonon propagators. The
finite Matsubara frequency phonons renormalize
the parameters a, b, and c
in the action, but it is the $\omega_m=0$
fluctuations that dominate the behavior as
$T\to T_c$. By treating a, b, and c
phenomenologically, we are able to run the
simulation on  substantially
larger lattices than previous quantum Monte
Carlo simulations of the
Holstein model.  This has allowed us a more
detailed look at the manner in
which the electronic properties are effected as
the Peierls-CDW transition is approached.

\acknowledgments

The calculations were carried out on the SGI Origin and IBM SP of the 
High Performance Computing Facility at the
University of Cambridge. DJS would like to
acknowledge support from the US Department of
Energy under Grant No.~DE-FG03-85ER45197.

\begin{figure}
\centerline{\epsfysize=6.00in
\epsfbox{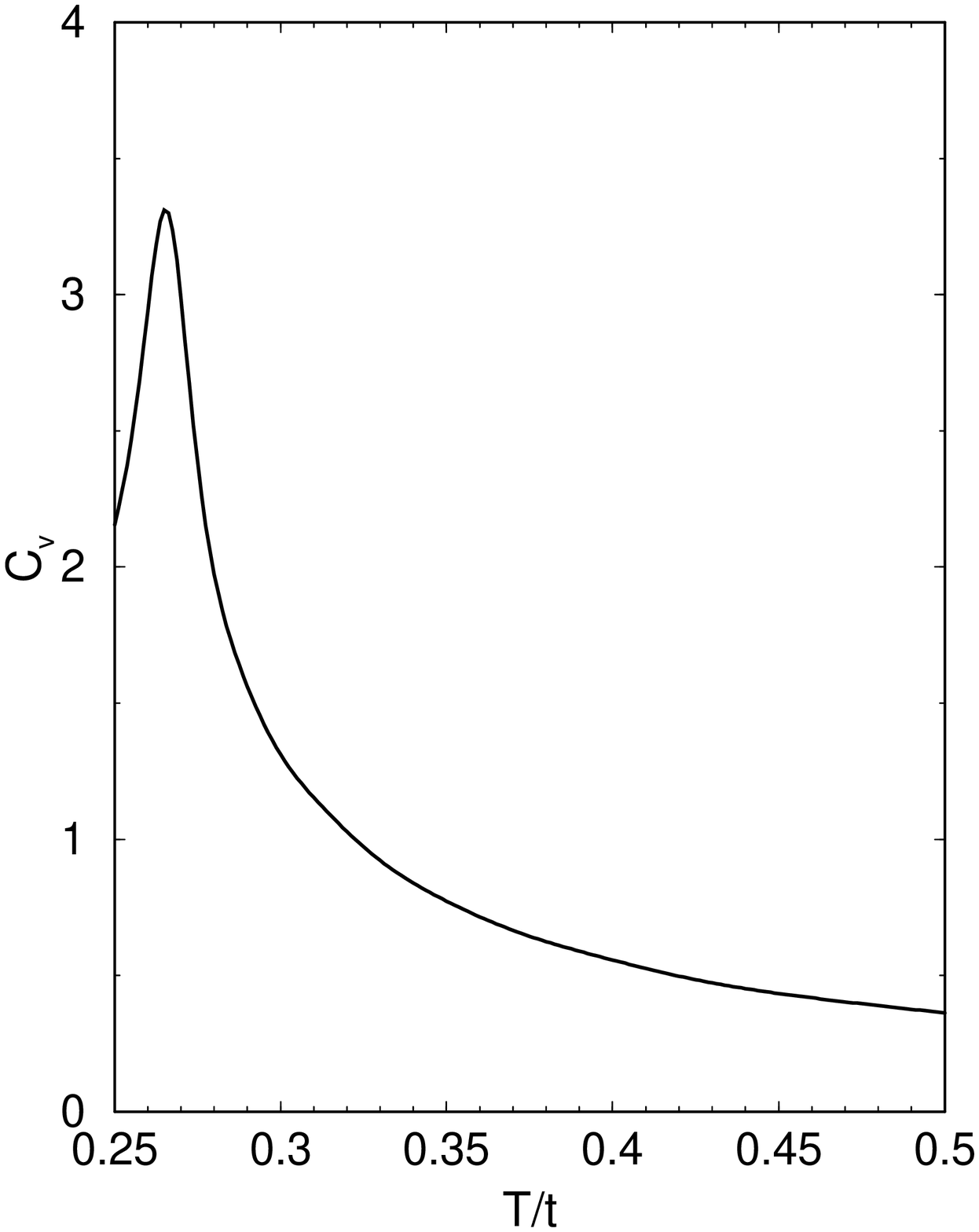}}
\caption{Monte Carlo results for the
Ising-specific heat versus $T$ for a $32\times
32$ lattice are shown for $a=1$, $b=2$, and
$c=1/8$.  All of our results are for these same
parameters. For this figure, we used 11 runs of 100,000 
measurements each. The statistical errors are of 
the order of the width of the line. }
\label{figone}
\end{figure}

\begin{figure}
\centerline{\epsfysize=6.00in
\epsfbox{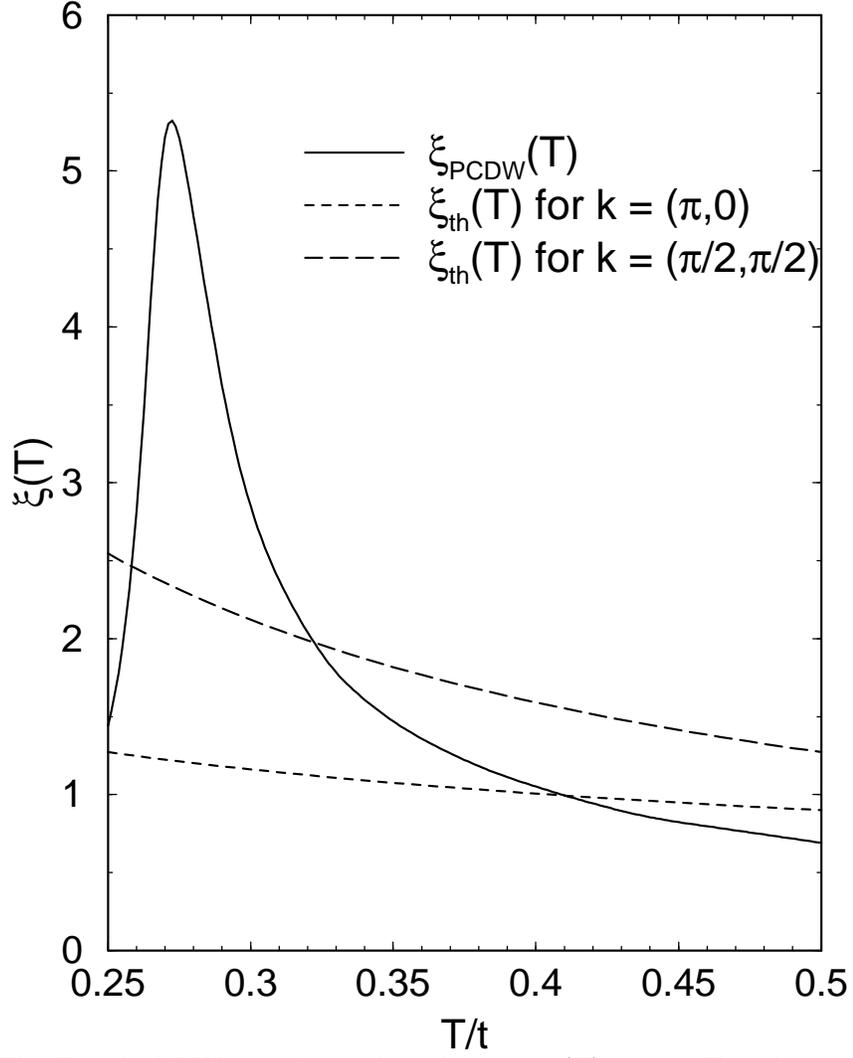}}
\caption{The Peierls-CDW correlation length
$\xi_{PCDW} (T)$ versus $T$ is shown as the solid
curve.  The non-interacting thermal,
quasi-particle, one-electron correlation lengths
for $k=(\pi/2, \pi/2)$ and $k=(\pi, 0)$ are shown as the long
and short dashed curves respectively.}
\label{figtwo}
\end{figure}

\begin{figure}
\centerline{\epsfysize=6.00in
\epsfbox{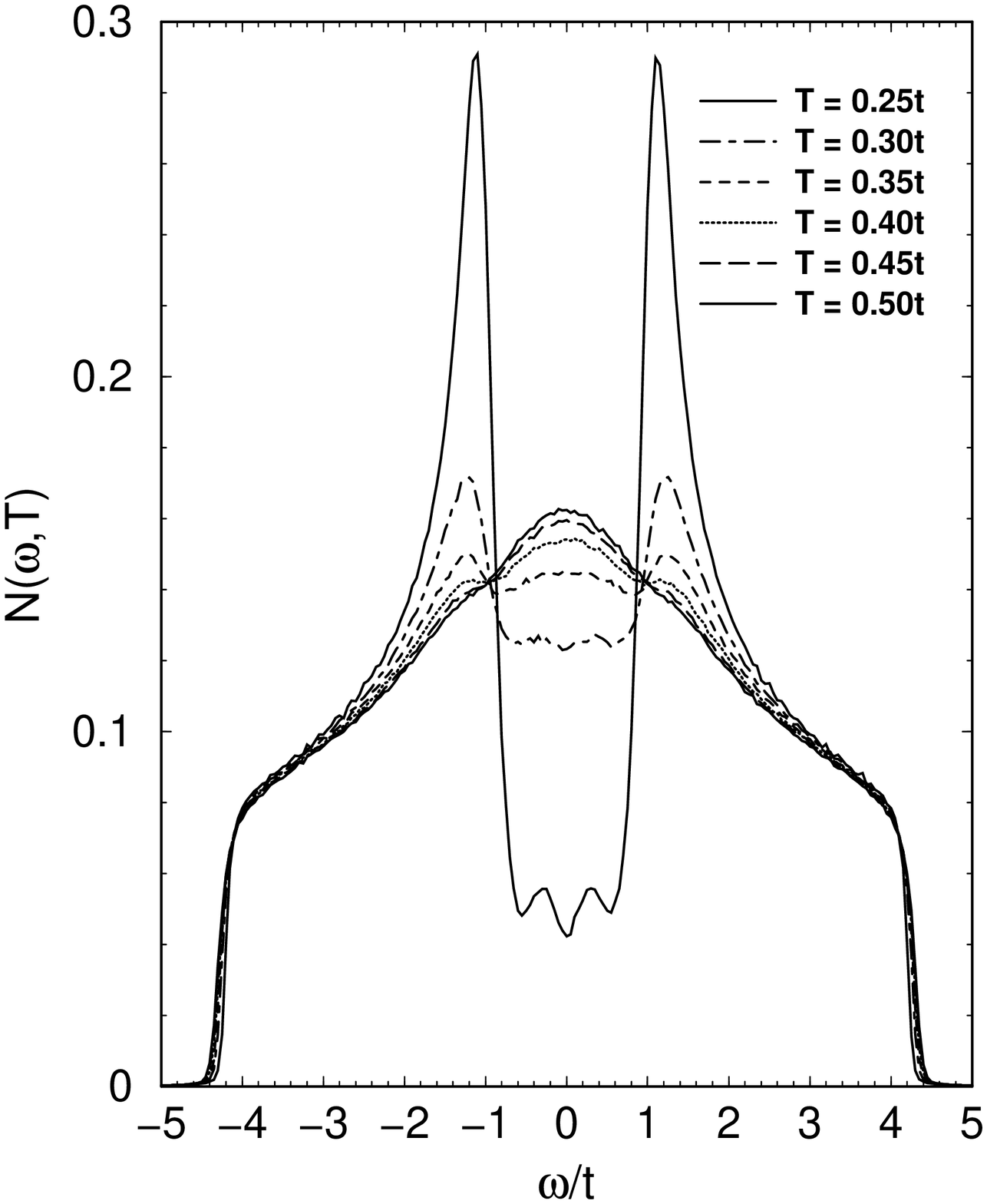}}
\caption{The single-particle density of states
$N(\omega)$ versus $\omega$ is shown for various
temperatures. the statistical errors (not shown) are
of the order of the linewidth}
\label{figthree}
\end{figure}

\begin{figure}
\centerline{\epsfysize=6.00in
\epsfbox{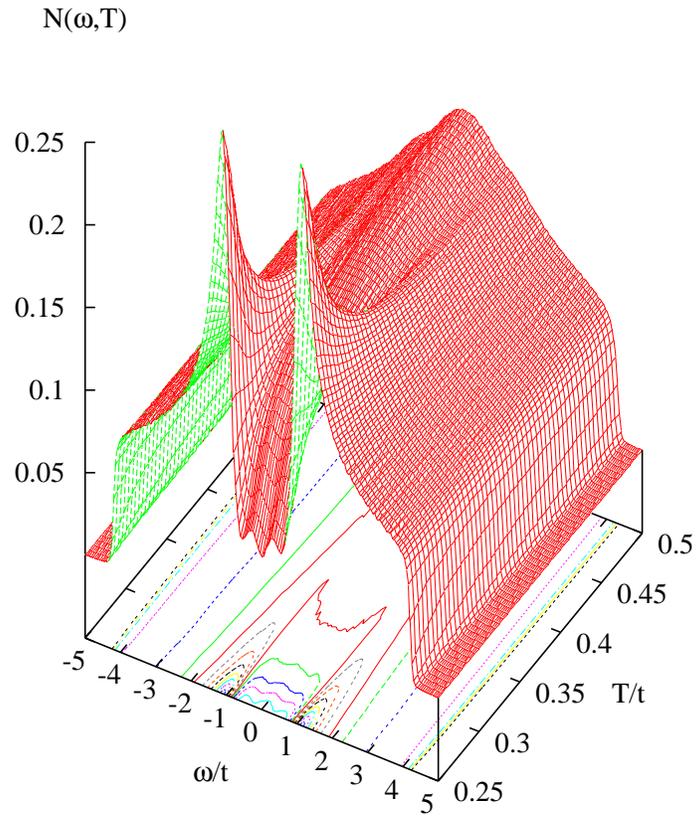}}
\caption{A plot of $N(\omega, T)$ versus
$\omega$ and $T$. The statistical errors 
which aren't shown for clarity are of 
the order of 0.3\%}
\label{figfour}
\end{figure}

\begin{figure}
\centerline{\epsfysize=6.00in
\epsfbox{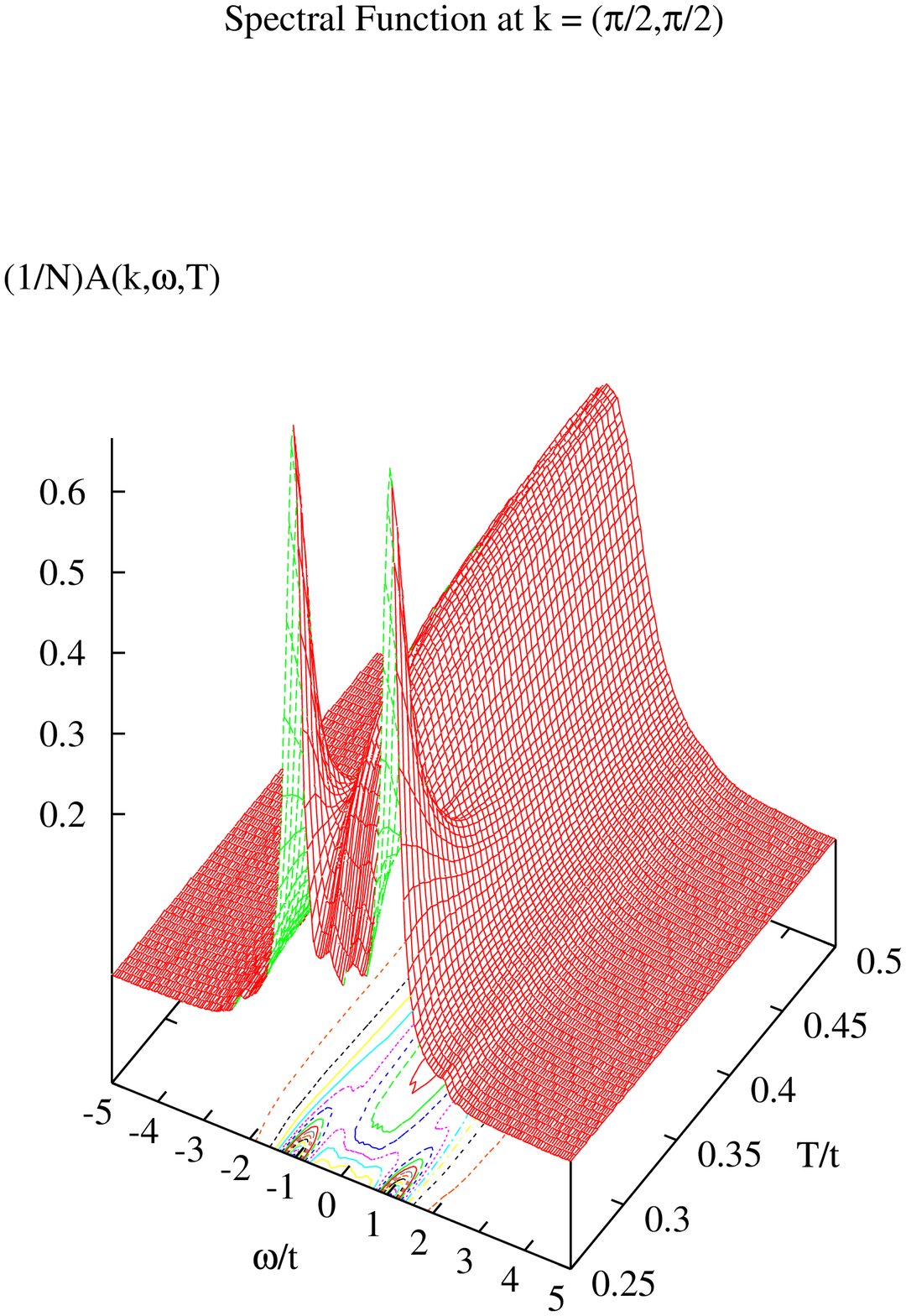}}
\centerline{\epsfysize=6.00in
\epsfbox{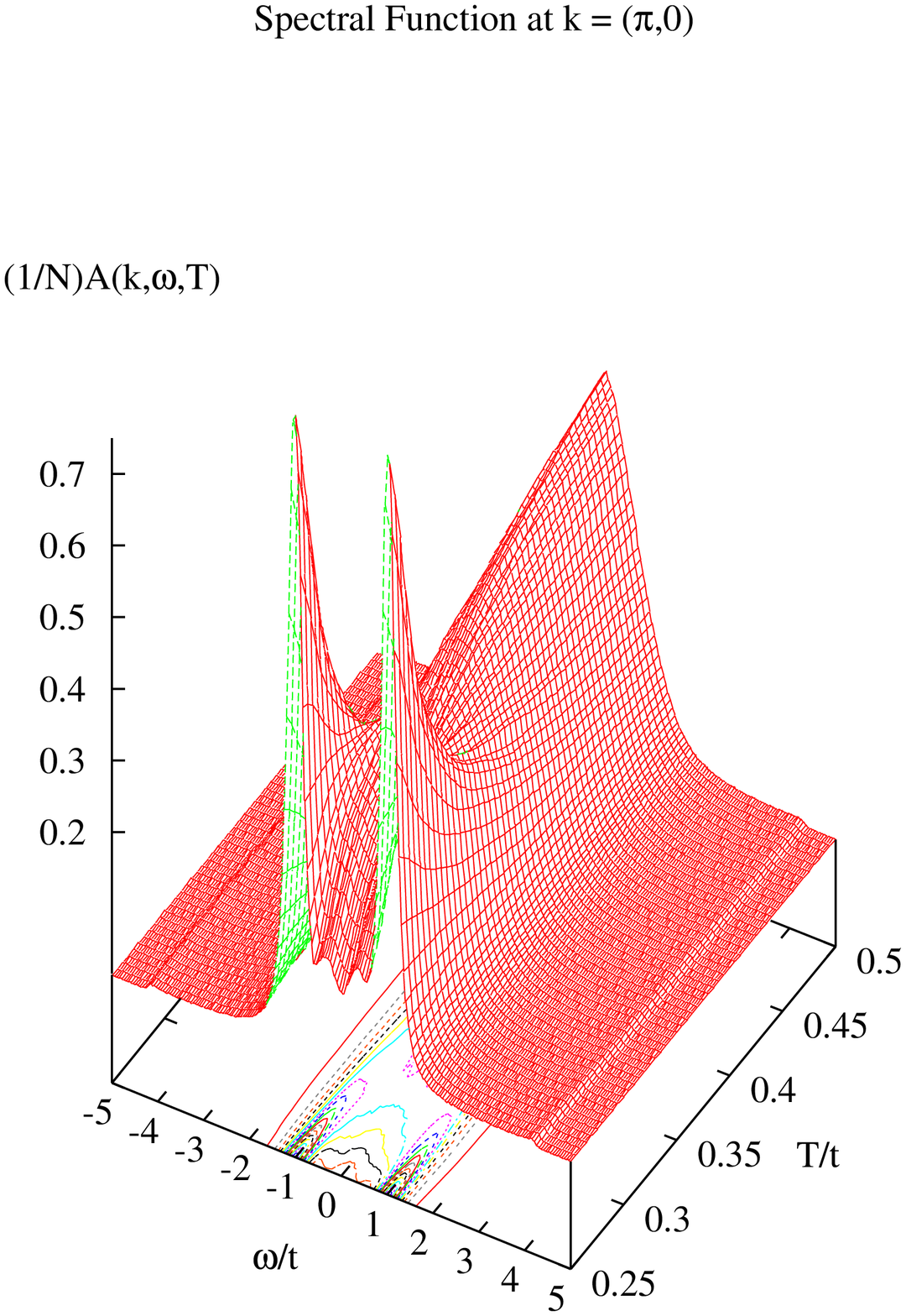}}
\caption{a) The single-particle spectral
weight $A(k, \omega)$ versus $\omega$ and $T$
for $k=(\pi/2, \pi/2)$ and b) for
$k=(\pi, 0)$. The statistical errors are 
of the order of 1\%}
\label{figfive}
\end{figure}

\begin{figure}
\centerline{\epsfysize=6.00in
\epsfbox{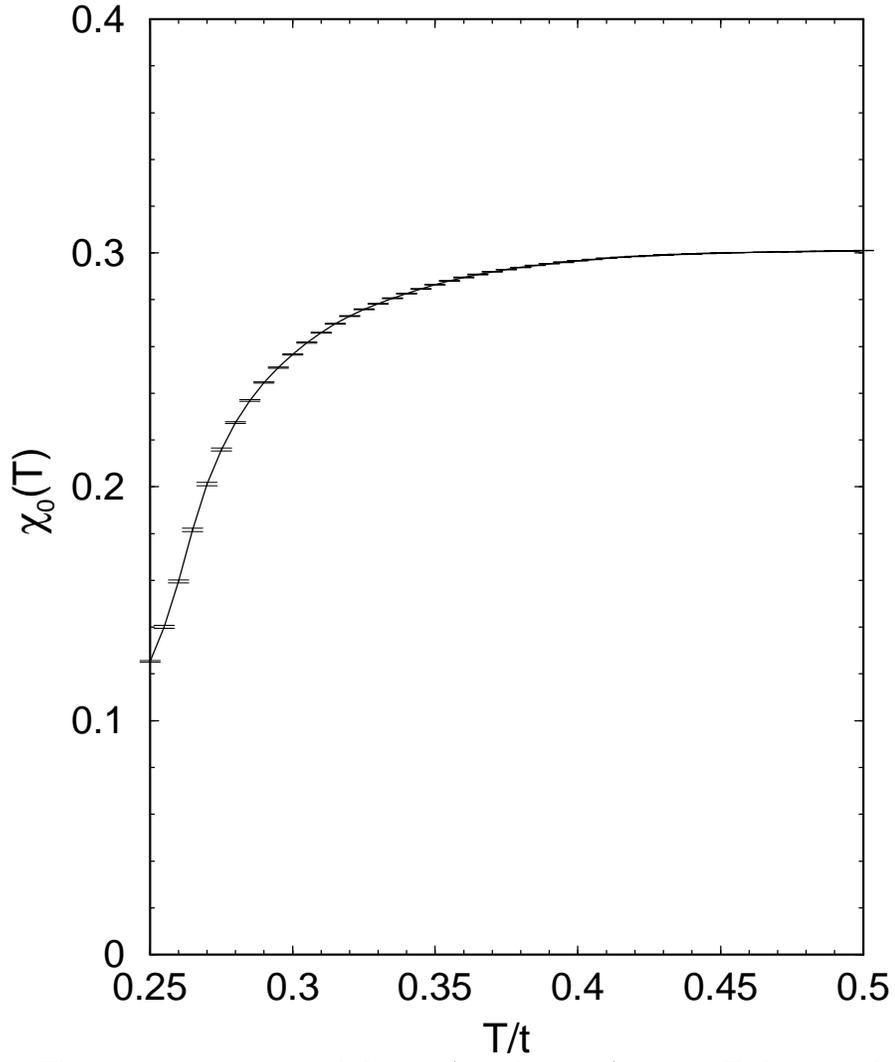}}
\caption{The static spin susceptibility
$\chi_0(q=0, \omega=0)$ versus $T$ showing the
opening of a gap as the temperature is
lowered.}

\label{figsix}
\end{figure}

\begin{figure}
\centerline{\epsfysize=6.00in
\epsfbox{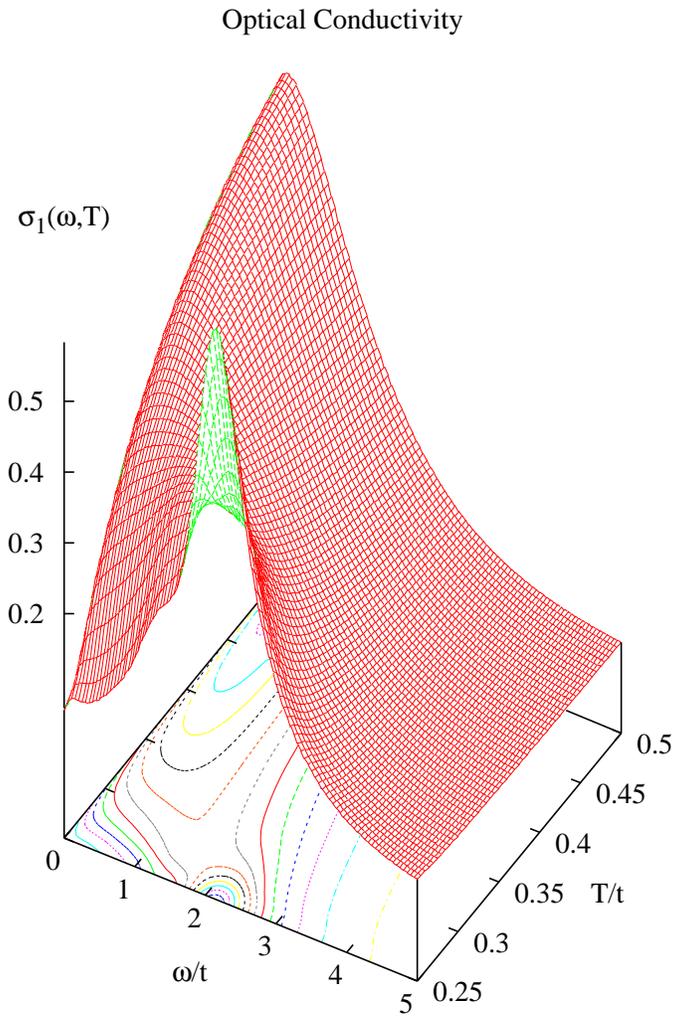}}
\caption{Plot of the real part of the optical
conductivity $\sigma_1 (\omega, T)$ versus
$\omega$ and $T$ showing the redistribution of
spectral weight to higher frequencies as $T_c$
is approached. The statistical errors, 
not shown, are of the order of 0.3\%}
\label{figseven}
\end{figure}


\begin{thebibliography}{99}
\bibitem{note} Note $\xi_{\rm PCDW}(T)$
decreases
below the ``ordering temperature'' and is
rounded
off by finite-size effects.
\end{thebibliography}
\end{document}